%% file: main.tex
\documentclass[11pt,a4paper]{article}
\usepackage{jinstpub}
\pdfoutput=1

\usepackage[colorinlistoftodos]{todonotes}
\usepackage[section]{placeins}
\usepackage[nottoc]{tocbibind}
\usepackage{caption}
\usepackage{subcaption}
\usepackage{lineno}
\title{Construction and testing of sMDT tubes
at the University of Michigan
for the ATLAS Muon Spectrometer upgrade 
}

\author[a, b]{C.~Wei}
\author[a]{A.~Chen}
\author[a]{D.~Amidei}
\author[a]{N.~Anderson}
\author[a]{E.~Carpenter}
\author[a]{L.~Cooperrider}
\author[a]{T.~Dai}
\author[a]{E.~Diehl}
\author[a,1]{C.~Ferretti}
\author[a, c] {J.~Li}
\author[a]{P.~Lugato}
\author[a]{J.~Minnella}
\author[a]{S.~Moskaitis}
\author[a]{K.~Nelson}
\author[a]{V.~Pillsbury}
\author[a]{E.~Salzer}
\author[a]{L.~Simpson}
\author[a]{Z.C.~Wang}
\author[a]{Z.R.~Wang}
\author[a]{C.~Weaverdyck}
\author[a, b]{Z.~Yang}
\author[a]{B. Zhou}
\affiliation[a]{Department of Physics, University of Michigan, 450 Church Street, Ann Arbor, MI 48109,  USA}
\affiliation[b]{Modern Physics Department, University of Science and Technology (USTC), Hefei, China}
\affiliation[c]{Physics Department, Shanghai Jiao-Tong University, Shanghai, China}
\emailAdd{ claudio.ferretti@cern.ch, Contact Editor}
\abstract{

This paper reports on the design and construction of infrastructure and test stations for small-diameter monitored drift tube (sMDT) assembly and testing at the University of Michigan (UM) to prepare for the ATLAS Muon Spectrometer upgrade for the high-luminosity program of the Large Hadron Collider.
Procedures of the tube assembly and quality assurance and control (QA/QC) tests are described in detail. 
More than 99\% of the tubes meet the tube QA/QC specifications based on 2100 tubes built at UM.
The UM test stations are also used for QA/QC testing on the tubes constructed at Michigan State University. 
These tubes are being used to construct the sMDT chambers which will  replace the current MDT chambers of the barrel inner station of the Muon Spectrometer.
 
}

\begin{document}

\maketitle
\flushbottom

%%%%%%%%%%%%%%%%%%%%%%%%%%%%%%%%%%

\input{overview.tex}

\input{infrastructure}

%  This tube_bending section should become a subsection of tube_construction    
%\input{tube_bending.tex}

\input{tube_construction.tex}
%subsections: Tube_Straightness, Tube%EndplugPrep, Wiring, swaging, leak_measurement

%rename visual_inspection to Quality Control
%subsections: visualinspection, length&tension, HV
%HV subsubsections: mounting, gas, power, doing the test, dealing with DC

\input{quality_control.tex}

\input{tube_length_wire_tension.tex}

\input{high_voltage_test.tex}

\input{tube_database.tex}

\input{conclusion.tex}

\newpage

\newpage

\bibliographystyle{JHEP}
\bibliography{bibliography}

\end{document}

%% file: overview.tex
\section{Introduction}

The ATLAS Muon Spectrometer will receive a major upgrade in  the next 
long shutdown (LS3) of the Large Hadron Collider (LHC) to cope with the
the much higher interaction rate of the  High-Luminosity LHC
(HL-LHC)~\cite{phase2TDRMuonSpectrometer}.
The central (barrel) region inner station of the Monitored 
Drift Tube (MDT) chambers will be replaced by new chambers using
small-diameter MDT (sMDT) tubes of 15 mm diameter, half that of the 
previous MDT. 
The smaller tube radius reduces both the maximum drift time (from 750~ns down to 180~ns) 
and the tube cross-section which lowers the tube occupancy by a factor 8 for improved performance in the HL-LHC.
Furthermore the smaller volume of the sMDT detector makes space for 
the installation of a new trigger tRPC (thin resistive plate chamber) detectors to improve the Level-1 muon trigger efficiency in the HL-LHC
environment.
\par
A total of 96 new sMDT chambers will be built to replace 
the Muon Spectrometer BIS1 to BIS6 MDT chambers, where BIS stands 
for Barrel-Inner-Small (sectors).
Half of these will be constructed at the Max Plank Institute for 
Physics in  Munich (MPI), Germany, and the other half in the US
at the University of Michigan (UM) and at Michigan State University 
(MSU).
The sMDT drift tube design and parameters are described in 
detail in~\cite{smdtDesign}.
Table~\ref{tab:parameters} lists the major tube parameters.
    \begin{table}[h]
        \centering
        \begin{tabular}{l|l}
             \hline
            Parameter & sMDT \\
            \hline
            Tube material & Aluminium AW6060-T6/AlMgSi \\
            Tube surface & Surtec 650 chromatization \\
            Tube outer diameter & 15.000 mm \\
            Tube wall thickness & 0.4 mm \\
            Tube length & 1615 mm \\
            Tube straightness & 0.5 mm / tube\\
            Wire material & W-Re (97:3) \\
            Wire diameter &  $\rm 50\,\mu m$ \\
            Wire resistance & 44 $\Omega$/m \\
            Wire tension & 350 $ {} \pm 15 \; $  g \\
            Gas mixture & Ar:CO2 (93:7) \\
            Gas pressure & 3 bar (abs.) \\
            Gas leak rate limit & $< ~ 1\times 10^{-8}~\frac{mbar ~\times ~cm^3}{sec}$ per tube\\
            Gas gain & $\rm 2\times10^{4}$ \\
            Wire potential & 2730 V \\
%            Maximum drift time & 180 ns \\
%            Wire positioning accuracy & $\rm 10\,\mu m$ r.m.s. \\
%            Wire dark current limit & $<$ 2 nA \\
            \hline
        \end{tabular}
        \caption{sMDT tube materials and operating parameters}
        \label{tab:parameters} 
    \end{table}
\par
\begin{figure}[hbt]
    \centering
              \subfloat[]{
    \begin{subfigure}{0.6\textwidth}
        \centering
        \includegraphics[width=0.99\linewidth]{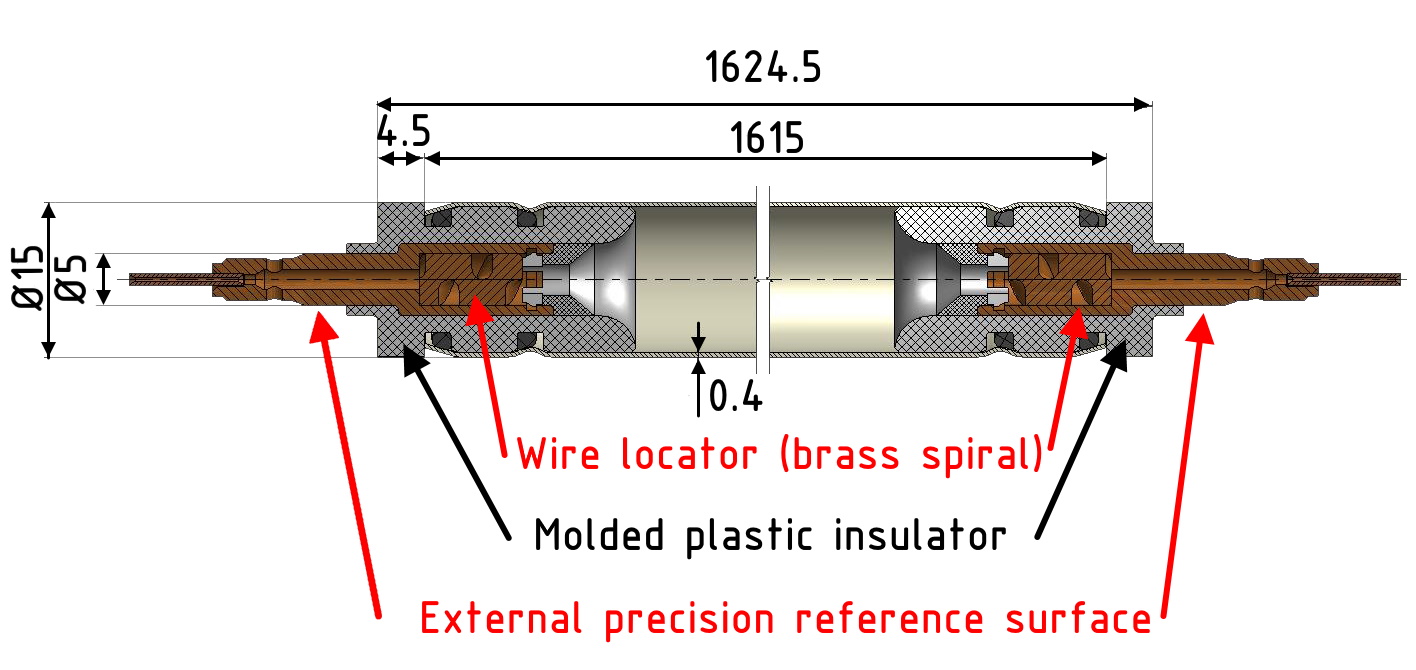}
    \end{subfigure}
    }
              \subfloat[]{
    \begin{subfigure}{0.39\textwidth}
        \centering
        \includegraphics[width=0.99\linewidth]{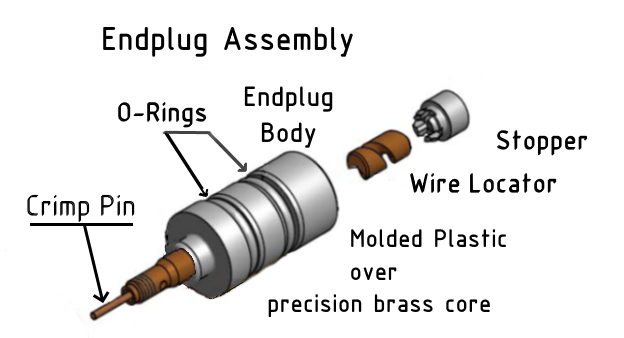}
    \end{subfigure}
    }
    \caption{(a) Tube design, with an exploded view of the end-plug components~\cite{smdtDesign}.  (b) The components and assembly of an endplug. All dimensions are in mm.}
    \label{fig:endplug}
\end{figure}

The endplugs  (Figure~\ref{fig:endplug})
consist of a precision brass wire locator, called a twister, inside a central brass insert (both manufactured by Poschl\footnote{Pöschl Präzisionsteile GmbH,
https://www.poeschl-gmbh.eu/home.html, Industriestr. 10, 82110 Germering Germany, Tel. 089/894454-0, Fax 089/89445444})
which is molded into a PBTP plastic body
manufactured by the Institute for High-Energy Physics (IHEP) at the Kurchatov Institute in Protvino, Russia. 
IHEP also makes the crimp tubes and stoppers (which 
lock the locators in the endplugs). 
Two grooves on the plastic outer surface hold o-rings 
which make a gas tight seal between the endplug and 
the tube inner wall.

\par
26000 precision tubes are needed for the Michigan chamber production.  
MSU is responsible for tube assembly and initial Quality Assurance (QA)
tests after which tubes are delivered to UM.
Tubes are tested again at UM to verify all
specifications are met before chamber assembly.
\par
Early in the R\&D period for sMDT construction tooling and the production cycle UM also built and tested 2600 tubes, 500 for R\&D in 2018 and 2105 for chamber mass production during the summer of 2021.
With three students working 8 hours per day, 
a production rate of 50 tubes/day was reached 
consistently in summer 2021, including 
all the QA tests on the constructed tubes. 
\par
This document describes the infrastructure and tooling used in 
building  tubes at UM, and the methods used to assess the
quality of all tubes used in sMDT chamber production at Michigan.
%\par
%\begin{itemize}
%    \item Visual inspection
%    \item Tube bending
%    \item Length measurement
%    \item Wire tension
%    \item Dark current measurement
%\end{itemize}

%% file: infrastructure.tex
\section{Infrastructure}

The ATLAS ``tube room`` is a $12\times 8 ~m^2$ lab space 
dedicated to drift tube assembly and QA testing. 
It is organized into multiple work stations for the construction and testing 
of sMDT tubes.
\begin{figure}[tbp]
    \centering
            \subfloat[]{
    \includegraphics[width=0.475\linewidth]{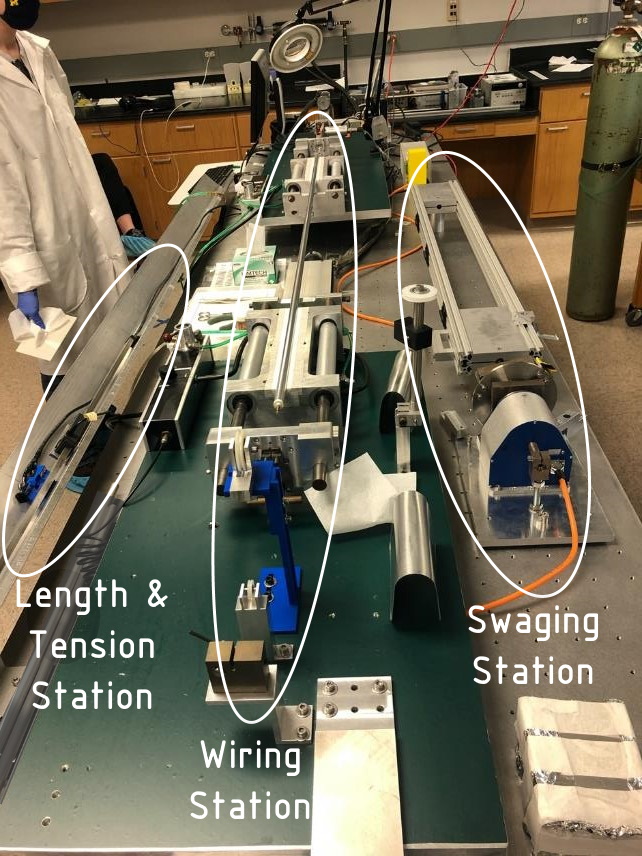}
    }
            \subfloat[]{
    \includegraphics[width=0.50\linewidth]{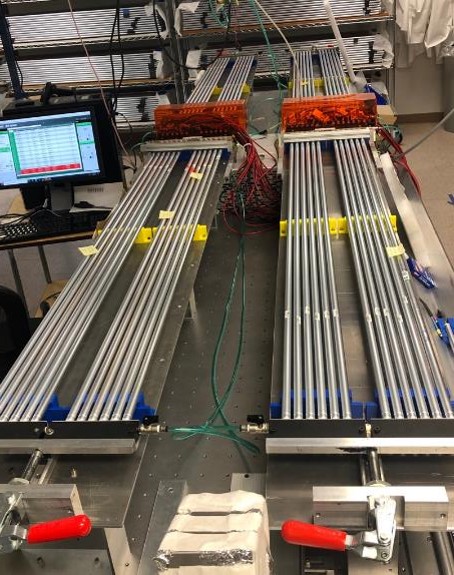}
            }
    \caption{(a) Tube construction stations: wiring and crimping (middle), endplug swaging (right), and the QA station for tube length and wire tension measurements (left). 
     (b) Tube QA dark current test station to test up to 48 individual tubes in parallel using a CAEN HV system.}
    \label{fig:wire-station}
\end{figure}
Figure~\ref{fig:wire-station}(a) and (b) show the tube construction station and the dark current test station in the tube room, respectively. Each station is set up on a $3.66\times 1.22~m^2$ flat optical table.
Once a tube is assembled, its straightness, gas tightness, 
length,  wire tension, and  dark current are 
measured. 
In addition, a large storage structure accommodating up to 3,000 tubes 
was built to handle the logistics for chamber mass production.
Table~\ref{tab:stations} lists the UM tube room work stations.
These stations are arranged so that all of them can 
be operated at the same time, optimizing the tube construction and test efficiencies.

\begin{table}[h]
\centering
    \begin{tabular}{l|l|l}
         Station name in the tube room & Purpose & Tubes per station\\
        \hline
         Wire stringing and tensioning & construction & single\\
         Swaging endplug onto tube & construction & single\\
         Length and tension measurement & QA & single\\
         Straightness measurement & QA & single\\
         Gas leak at 3 barA measurement & QA & single\\
         Dark current measurement & QA & 48\\
         Reverse HV treatment & QA & 48\\
        \hline
    \end{tabular}
    \caption{UM tube room stations. }
    \label{tab:stations} 
\end{table}

Lab cleaning is performed twice a day to maintain
a pristine environment.
Personal protection equipment (lab-coat, shoe covers, 
hair nets, gloves and disposable masks) are easily
accessible immediately upon entry and are required 
at all times, with the exception of the gloves which are used 
only when handling tubes.

%% file: tube_construction.tex
\section{Tube Construction}

%Michigan State University (MSU) did the assembly of almost all of the 26000 tubes needed for Michigan production of 48 chambers. 

The tube construction infrastructure and test 
stations were designed and built at UM during the 
R\&D phase (2018-2020) of the ATLAS muon detector 
upgrade project. Using raw tubes from the 
manufacturer Mifa Alumninum\footnote{Mifa Aluminium, Rijnaakkade 6, 5928 PT Venlo, E, The Netherlands: https://mifa.eu/en, info@mifa.nl, T: +31 77 389 88 88, F: +31 77 389 89 89}
500 drift tubes were constructed and 
tested to commission and verify the 
infrastructure and tooling. These tubes were used 
to build an sMDT prototype chamber at UM 
to demonstrate the tube construction and test 
procedures. Michigan sMDT chamber mass production 
started in May 2021. 
Knowledge and technologies for tube production 
and testing were shared with MSU for tube mass 
production. 
One batch of 2105 sMDT  tubes was built at the UM 
site in the summer of 2021 to supplement the tube 
construction effort at MSU to assure that the 
early sMDT chamber production schedule at UM 
could be met. 

The tube construction process involves three 
main steps: preparation of raw aluminum 
tubes; tube assembly; and
quality assurance tests of each constructed tube.
This section will focus on the construction 
process carried out at the UM site.

\subsection{Tube straightness measurement} \label{sec:TubeBendingMeasurement}

Mifa supplied all raw sMDT tubes for both MPI and 
Michigan, delivered in crates of 702 tubes each.
Michigan's tubes were sent by truck to CERN and 
then by air freight to Michigan.  
All incoming tubes were inspected at CERN and 
at UM before tube construction.
%Early measurements of the diameter and of the %cylindricity were well within specifications.
\par
\begin{figure}[ht]
    \centering
            \subfloat[]{
    \includegraphics[width=0.31\linewidth]{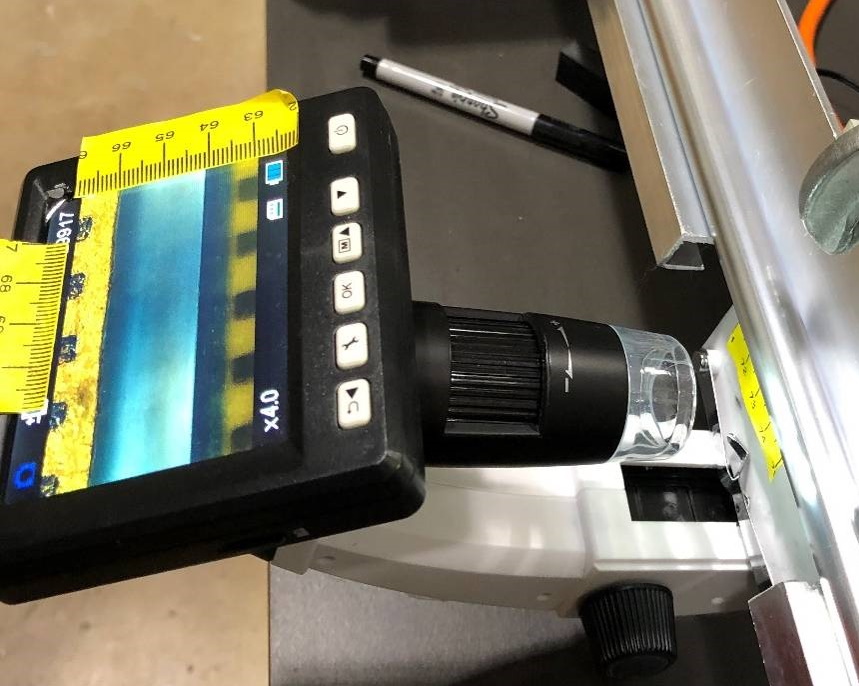}
    }
            \subfloat[]{
    \includegraphics[width=0.34\linewidth]{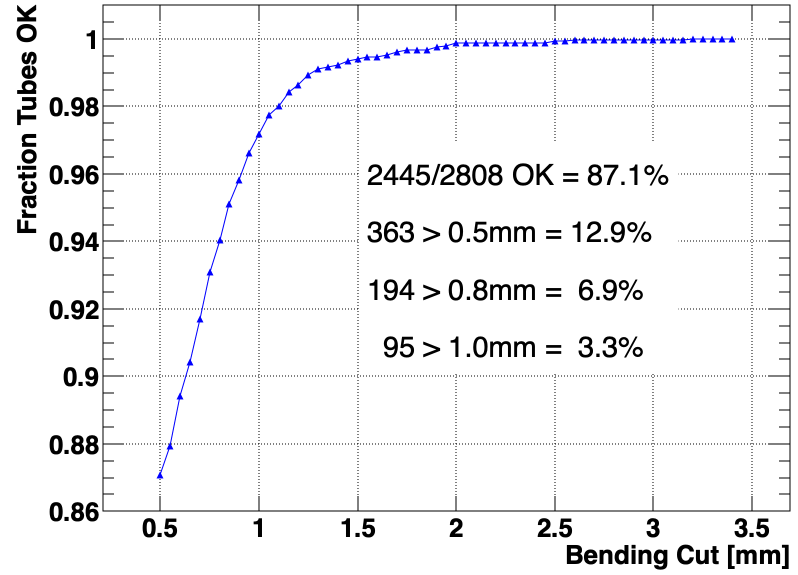}
    }
            \subfloat[]{
    \includegraphics[width=0.34\linewidth]{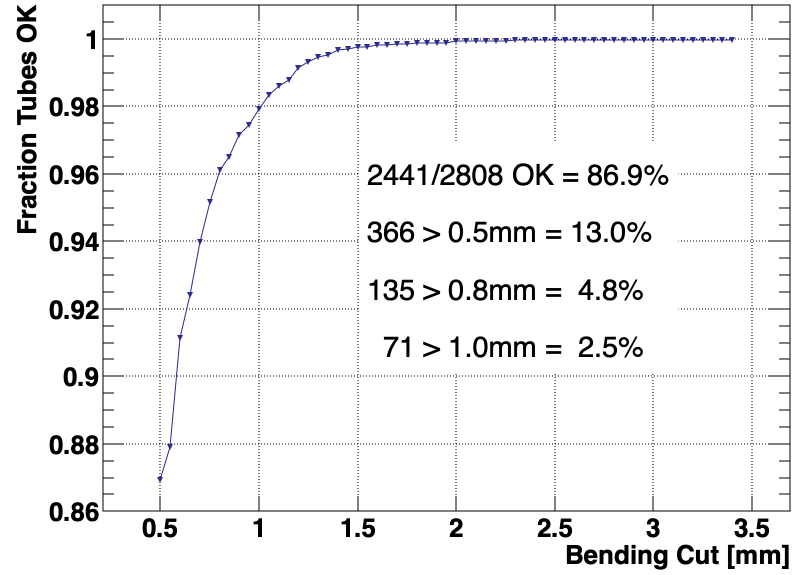}
    }
    \caption{Straightness test station (a) and results of straightness measurements, the fraction of tubes accepted depending on a bending value, of raw tubes delivered in 2020 (b) and 2021 (c).}
    \label{fig:Bending}
\end{figure}

One requirement is the tube 
straightness: 
the specification calls for no more than 0.5~mm 
deviation when a tube is resting on a flat
surface.
%Though curved tubes will be forced into relative
%straightness by the jigging used when gluing the 
%chamber, 
%they will still apply some force on their 
%immediate neighbors and build internal stresses %into the finished chamber.
However, nearly $ 20\%$ of early 
tube shipments exceeded this specification.
To measure tube straightness a system with an 
optical technique using a microscope was 
developed. 
The system consists of a straight V-shape bar that has 
a flat surface (flatness < 0.05~mm) with straight 
backstop to keep the center-line of the tube at 
the focal point of the microscope. 
The microscope looks at the the center bottom of 
the tube where it should touch the flat surface. 
The tube is rotated in the V bar to allow measurement of the maximum distance 
between the tube and the surface.
The microscope has a built-in screen which is
set to 20X magnification (see picture on 
Figure~\ref{fig:Bending} (a)).
This technique assumes that the greatest out-of-
straightness deviation (refereed to as sagitta) 
will be at the center along the tube length.
Plots in Figure~\ref{fig:Bending} (b) and (c) 
show the fraction of tubes accepted in two lots
as a function of the cut used to reject bent tubes. 
About 87\% of the tubes met the specification 
of the tube straightness requirement.
Based on these measurements it was decided to 
relax the original cutoff specification and to 
only reject tubes with more than 0.8~mm sagitta, 
about 6\% of those delivered in 2020 and 2021.
The rejected tubes (slightly over 1,000) 
were sent back to the manufacturer for 
straightening treatment and returned to Michigan 
at the end of 2021.
Three identical optical measurement devices were 
built at UM and sent to the tube production 
company (MIFA in Netherlands), CERN, and MSU. 
This test is now a routine check of the tube
mass production.
After using the UM measurement technique the 
company adopted a stricter
QA/QC protocol which reduced the 
fraction of bent tubes delivered to about 2\%.

\subsection{Aluminum tube and endplug preparation}
Although tubes are cleaned at the factory, it was 
found that additional cleaning reduced the dark 
current drawn by tubes.
In the initial stage of R\&D tube production 
$\sim$12\% of tubes exceeded the dark current limit.   
To reduce this high dark current failure rate
the 2105 aluminum tubes used for summer 2021 tube production 
at UM were each re-cleaned by swabbing the 
interior with isopropyl alcohol (IPA) 
soaked wipes prior to the tube construction.
The interior of each tube was swept three times 
with fresh lint-free wipes wrapped around a small 
cylindrical piece of foam and soaked in IPA for 
each sweep.
This cleaning plug fit snugly inside the tube and 
was pushed through the tube with a clean acrylic 
rod to ensure the entire length of the interior 
was wiped clean.
The exterior was also cleaned with IPA-soaked 
lint-free wipes.
The cleaned tubes were stacked on foam saddles 
from the original packaging and the ends were 
lightly covered with lint-free wipes to allow 
the IPA to evaporate completely.
After all tubes were cleaned, the two ends of 
each tube were temporarily closed with clean 
rubber caps to keep the interior free of dust.
The thoroughly cleaned tubes had a failure rate
of only $\sim$3\% in the dark current test, significantly 
lower than that observed in early tube tests.
In addition, it was found that negative HV treatment,  described in 
Section~\ref{sec:DC-test}, could recover any tubes failing the initial dark current test.
%Tubes thoroughly cleaned and not performed substantially the same way in terms of dark current, but the statistics is limited since the IPA swabbing procedure was utilized for the vast majority of the tubes produced at UM.
\par
The endplug is assembled with two o-rings, a 
twister, and a stopper 
(see Figure~\ref{fig:endplug} (b)).
Individual parts were given a 15 minute 
ultrasonic bath in IPA before assembly 
and a further 5 minute ultrasonic bath after assembly.
After air drying the endplug assemblies were 
stored in clean plastic packaging until used.

\subsection{Tube assembly}
This section describes the tube construction 
stations for tube wiring and swaging, and the 
tube assembly procedures.

\subsubsection{Wiring station}

\begin{figure}[h]
%    \begin{subfigure}{0.5\textwidth}
%        \includegraphics[width=0.9\linewidth, angle=180.]{images/WireSpoolMagneticBrake.jpg}
%        \caption{Wire spool with brake.}
%        \label{fig:wire_spool_with_brake}
%    \end{subfigure}

    \begin{subfigure}{0.27\textwidth}
        \subfloat[]{
        \includegraphics[width=0.95\linewidth]{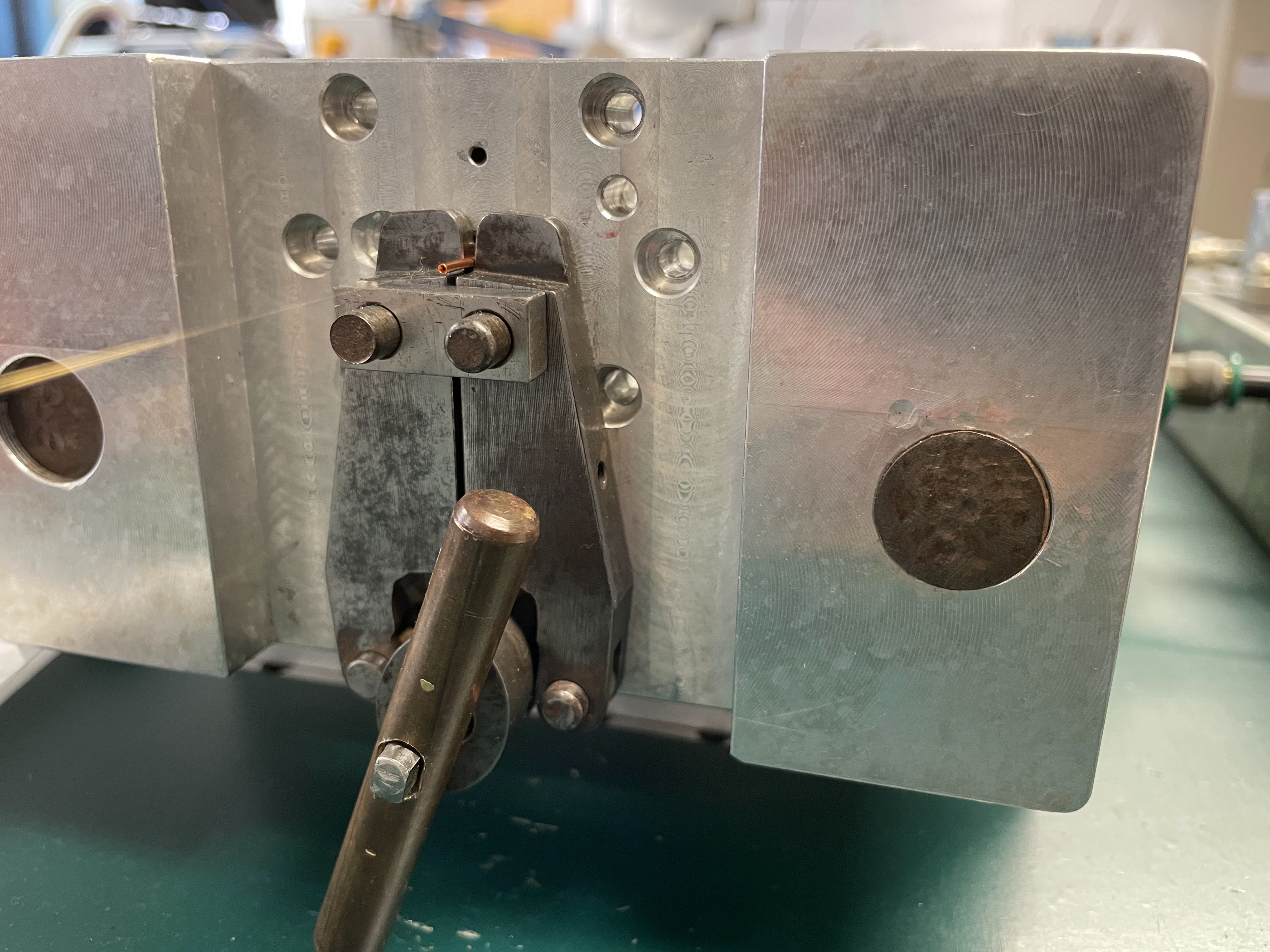}
        }
        \label{fig:crimp-jaws}
    \end{subfigure}
        \begin{subfigure}{0.72\textwidth}
        \subfloat[]{
        \includegraphics[width=0.95\linewidth]{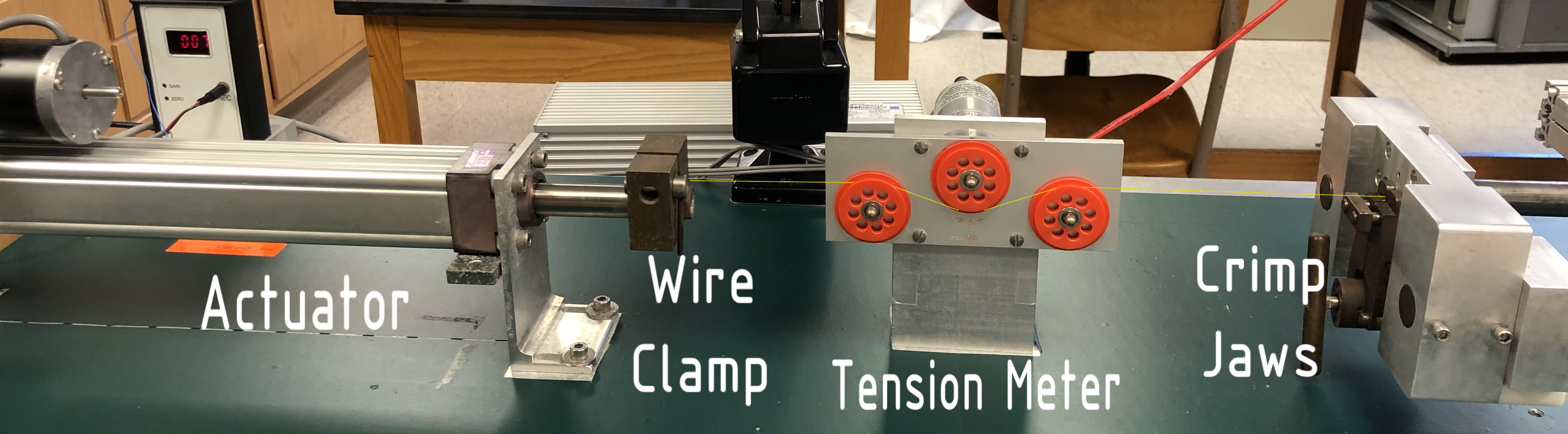}
        }
        \label{fig:actuator}
    \end{subfigure}
    \caption{(a) Wire crimp jaw. (b) The wire pulling actuator and tension meter used to stretch the wire before crimping. }
    \label{fig:tensioningtools}
\end{figure}

The tooling for wire stringing consists of two 
vacuum chucks that hold the tube rigidly in a
straight line and aligned with the wire feed.
Each chuck is fitted with a pneumatically
controlled end-plate on which is mounted a 
pair of cam-operated crimp jaws 
(Figure~\ref{fig:tensioningtools}a) which 
squeeze the 1.0~mm O.D. crimp pin down to 
0.7~mm.
The spool of wire\footnote{Luma Metall Fine Wire, Kalmar, Sweden} is mounted on a 
magnetically damped 
spindle~\footnote{Warner Electric Precision Tork MC2, 1-20 oz-in torque} to allow smooth
pulling.  
At the other end is a three wheel tension 
meter~\footnote{Electromatic Equipment Co., Tension Sensor model TE-500-22}
and a linear actuator~\footnote{Parker Linear Motion Drive Model No. ETS32-802QA21-GK200-A} 
for stretching the wire to the desired tension 
(Figure~\ref{fig:tensioningtools}b).

\subsubsection{Swaging station}

\begin{figure}[ht]
\centering
         \begin{subfigure}{0.45\textwidth}
        \centering
        \includegraphics[width=0.9\linewidth]{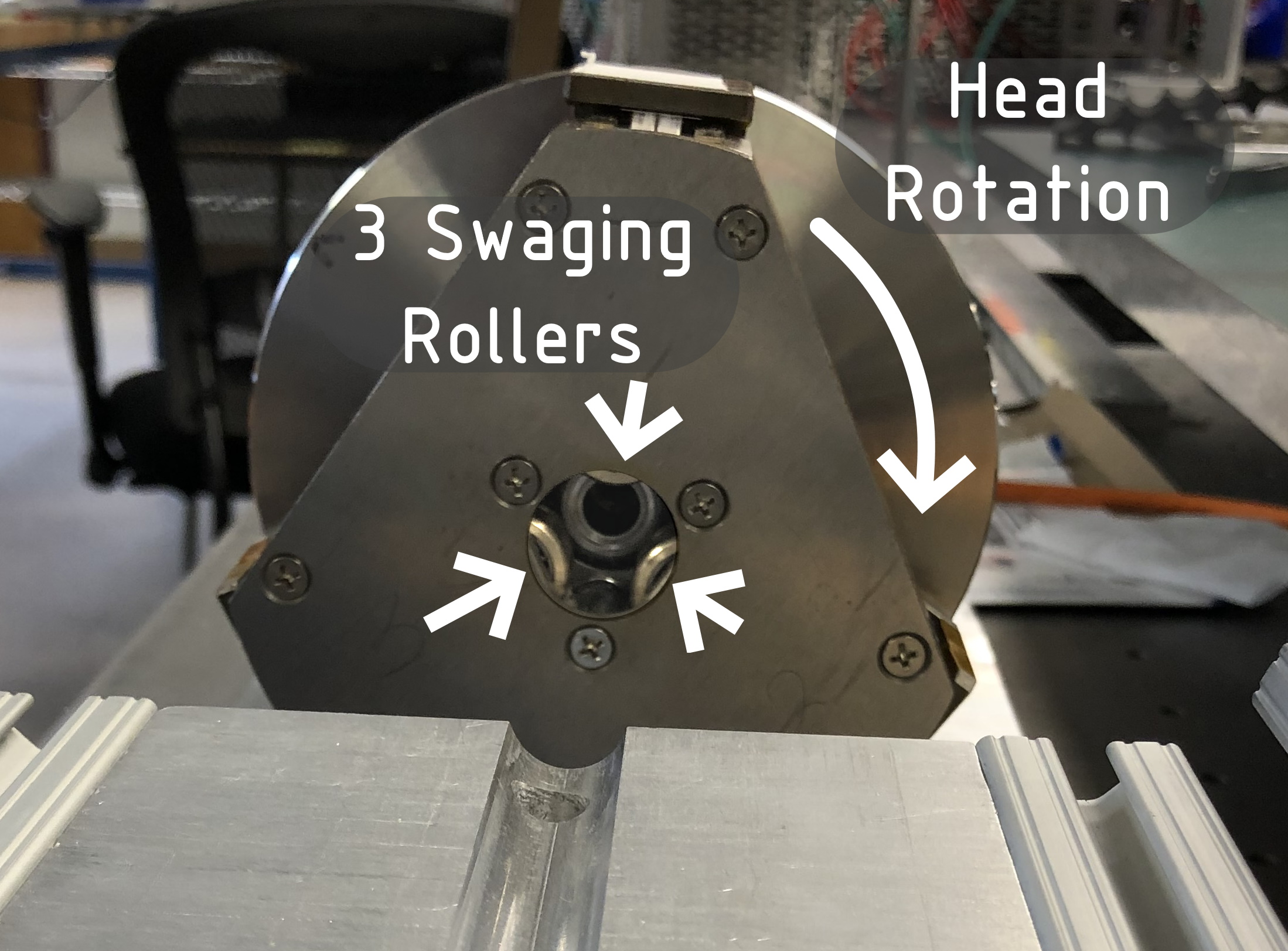}
        \caption{Swaging rotational head}
        \label{fig:swager}
    \end{subfigure} 
         \begin{subfigure}{0.45\textwidth}
        \centering
        \includegraphics[width=0.9\linewidth]{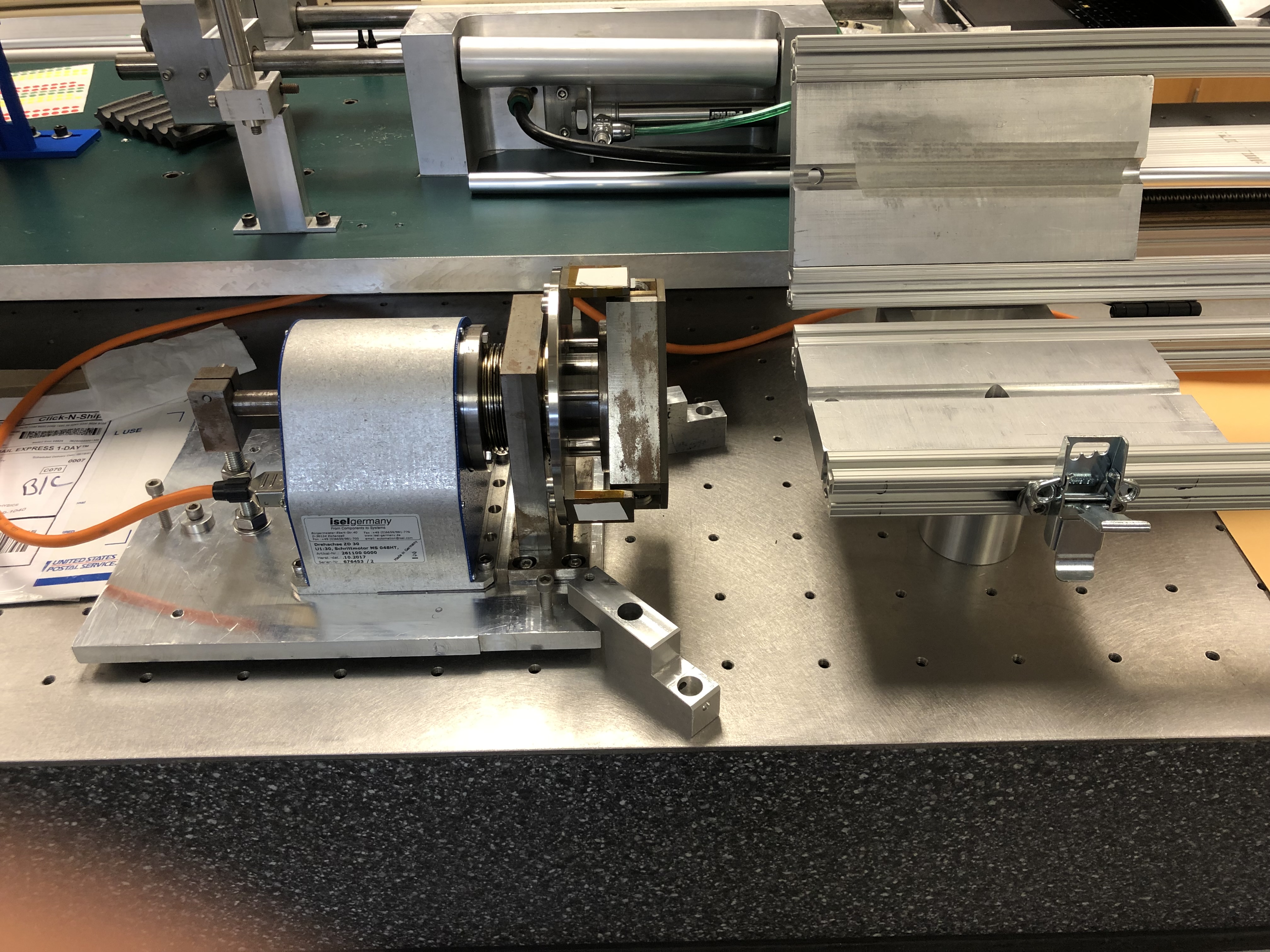}
        \caption{Side view of swaging set up}
        \label{fig:swager-side}
    \end{subfigure} 
    \caption{The rotary  swaging head for creating indents around the tube at the o-ring locations}
    \label{fig:crimpandswage}
\end{figure}

The second station used in tube construction is 
the swaging station 
(Figure~\ref{fig:crimpandswage}), where
the thin walls of the tube are deformed (swaged) 
to compress and seal against the o-rings on the
endplugs. 
The swaging head is a custom built rotary 
device that gradually forces 3 roller heads 
radially inward as they are rotated around 
the tube (Figure \ref{fig:swager}).
The head, designed at MPI and built at MSU,
is driven by an Isel rotary motor~\footnote{Isel Model Drehachse ZD 30, Belt drive with stepper motor, Model No. ETS32-802QA21-GK200-A}.
Figure~\ref{fig:swager-side} shows the side 
view of tube swaging set up,
including a structure to hold the tube when 
inserting it into a rotational head.

\subsubsection{Assembly procedure}

The tube construction process starts by placing an 
aluminum tube on the vacuum chuck to hold the tube securely.
The tube interior is vacuumed to remove any residual 
dust.
The $\rm 50\,\mu m$ tungsten-rhenium wire is 
secured to a shuttle which is carefully pulled
through the tube using a gentle vacuum. 
The wire is drawn over pulleys which prevent the wire from contacting the tube wall.
A magnetic brake controls the un-spooling wire 
to prevent tangles while wiring the tube.
The wire is then cut, leaving 0.5~m excess wire 
at each end.
The wire ends are threaded through
endplugs at each end of the tube and then the endplugs are 
inserted into the tube ends until seated.
Next, crimp pins are threaded onto each wire end and 
inserted into the endplugs.
One end of the wire is then secured to a 
stationary clamp.
The other end is fed through the tension monitor, 
stretched by hand to around 200~g and then 
secured to another clamp attached to the 
actuator~(Figure \ref{fig:tensioningtools}b). %\ref{fig:actuator}).
The fixed-end crimp pin (soft copper, 1~mm OD) 
is crimped (pinched by jaws to 0.7~mm, see 
Figure~\ref{fig:tensioningtools}a). %\ref{fig:crimp-jaws}). 
\par
An online control program based on LabView~\cite{bitter2006labview} 
moves the actuator and monitors the wire tension 
while the wire is gradually pulled until the 
tension reaches 400~g.
This tension is held for 30 seconds, and then 
the tension is slowly reduced to 325~g. 
Note that later in the process the tube swaging 
will increase the wire tension to 350~g.
The second crimp pin is pushed with tweezers 
into the endplug until it touches the brass 
core to insure the tension is not altered by the movement of the 
crimp pin.
The crimped thickness of each crimp pin is 
measured to ensure that it is no more than 0.71~mm which is
the value needed to hold the wire securely.
Excess wire protruding from the crimp-pins is 
cut off leaving a stub of about 2~mm.
\par
After the tube is wired, the wire tension is 
measured in the tension test station (details 
in Section~\ref{sec:WireTension}) 
to verify the tension value as set.
At this point, a tube with an insufficient wire 
tension can easily be disassembled and all 
components, except the wire and crimp-pins, 
can be salvaged for another construction attempt. 
%\subsection{Endplug swaging}

Each end of the wired tube is then mechanically 
deformed in the swaging station to lock the
endplugs in place and produce a tight o-ring
seal between the plastic endplug and the
interior wall of the tube.
This is done with the rotating set of 
three rounded rollers that create smooth indents 
around the circumference of the tube at each of
the two o-ring 
locations~(Figure~\ref{fig:swager}).
The innermost indent is made such that the width 
of the tube in the groove measures 13.7~mm 
or less.
This swaging process introduces an additional 
25~g of tension in the wire due to the slight
stretching of the tube.
After swaging a final tension test is done to 
confirm that the wire inside the constructed
tube has a tension of $350 \pm 20$~g.
Once the two endplugs are swaged the tube 
construction is done and a bar code sticker
is put on the tube with a unique serial
number for identification.

%% file: quality_control.tex
\section{Quality Control} \label{Quality Control}

Both individual parts and assembled tubes must pass a number of quality checks.
Random checks of the endcap brass precision brass surface are done on each batch received to ensure the diameter is within the specification of 
5.000$^{+0.00}_{-0.01}$~mm.
Tube roundness and diameter are verified with the precision 
endplugs to be well within tolerances. The straightness was 
discussed in section \ref {sec:TubeBendingMeasurement}.
\par
The QA tests on all constructed tubes, whether built at MSU or UM, include components visual inspection, tube straightness, 
leak tightness, assembled tube length, wire tension, and dark current.
The detailed procedures and test results are described in the
following sections.

\subsection{Tube visual inspection}

Constructed tubes are first given an visual inspection to
check for defects such as dents or cracks in the 
tube wall, badly swaged endplugs, or severely bent crimp pins.
Bent crimp pins are carefully straightened.
Tubes with unrecoverable problems are rejected and flagged in 
the database.

\subsection{Tube leak tightness test}
\begin{figure}[htb]
    \centering
            \subfloat[]{
    \includegraphics[width=0.33\linewidth]{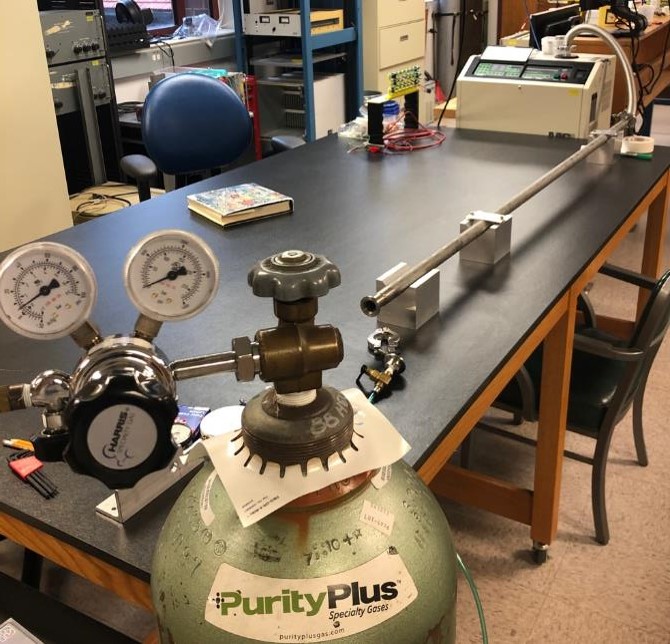}
    }
            \subfloat[]{
    \includegraphics[width=0.32\linewidth]{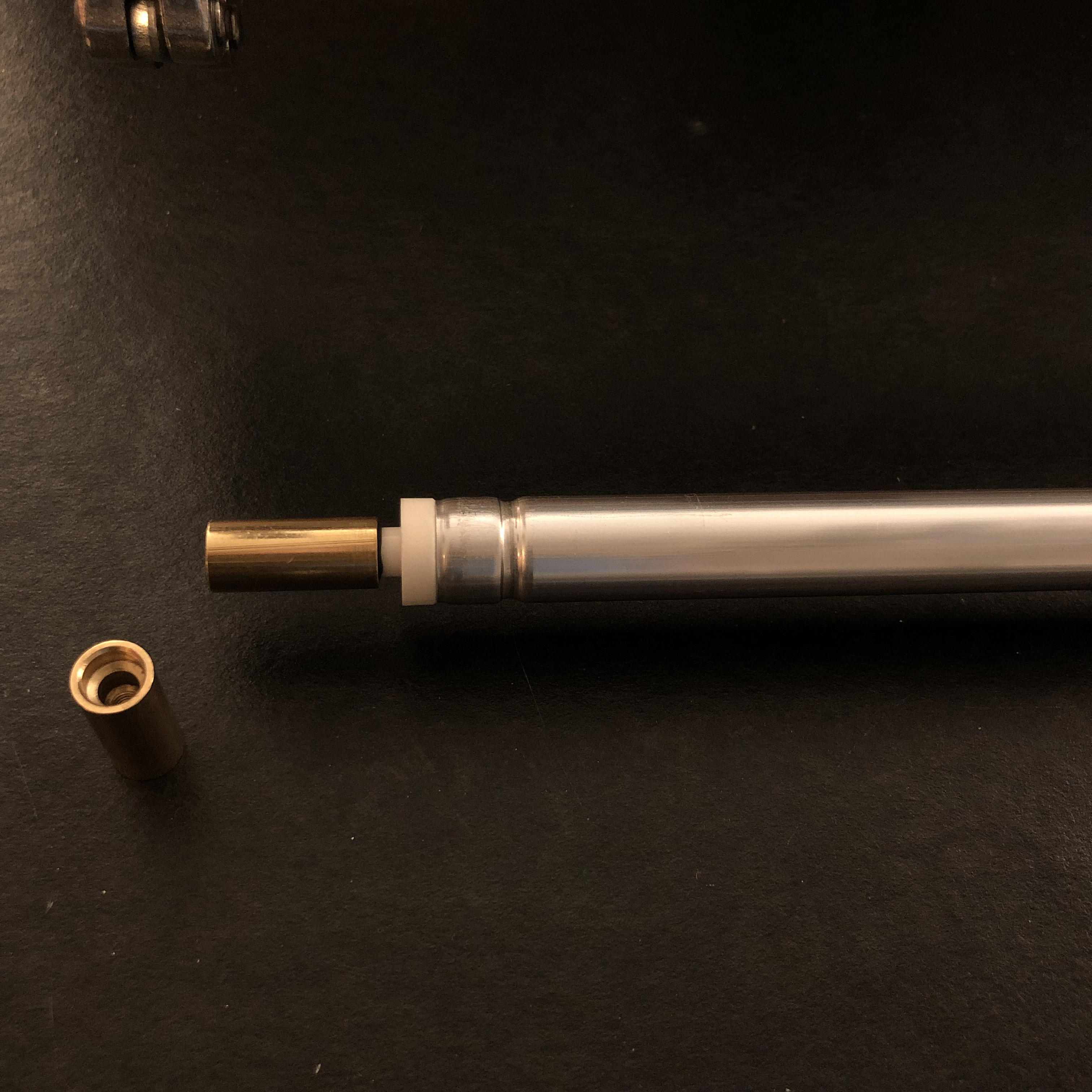}
    }
            \subfloat[]{
    \includegraphics[width=0.32\linewidth]{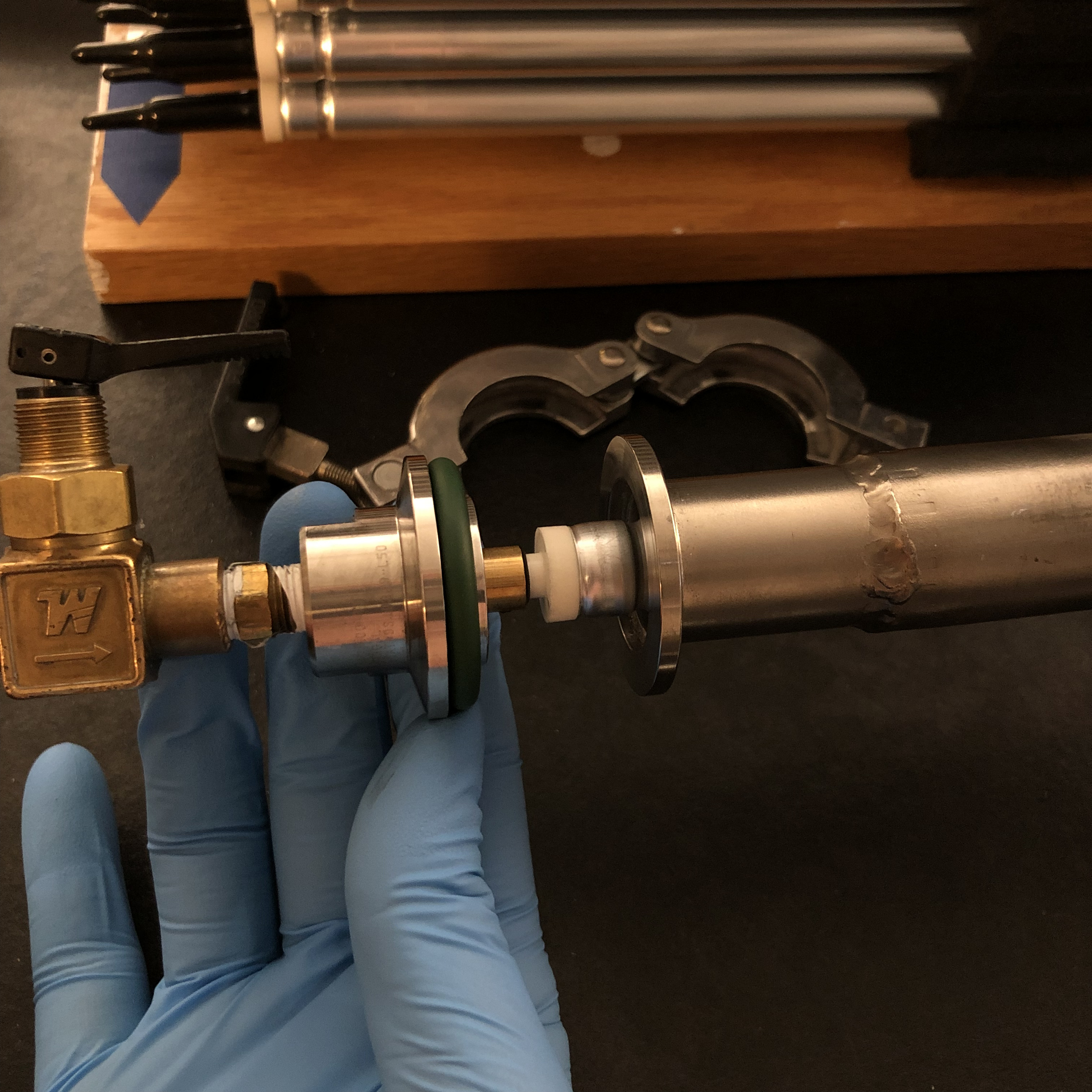}
    }
    \caption{(a) Set up of the tube leak test station.
    (b) Interior end of the tube is sealed with this 
    brass piece.
    (c) Feed-through end of the vacuum vessel with a 
    tube connected. Most of the tube is inside the vacuum vessel. 
    The tube is pressurized with Helium to 3 barA.
    }
    \label{fig:leaktest}
\end{figure}

The constructed tube is required to be leak tight with an upper limit of 
$\rm 1\times10^{-8} cm^3\times mbar/s$ with the tube at 3 bar (absolute) pressure. As shown in Figure~\ref{fig:leaktest}(a),
the leak rate test is performed with a helium leak detector connected to a vacuum vessel 25~mm ID x 1.8~m long. 
One end of the tube being tested is sealed with a brass cap 
and an o-ring (as shown in Figure~\ref{fig:leaktest}(b), the other 
is screwed into the cap of the vacuum vessel, which allows the tube
to be pressurized with 3 bar Helium (see Figure~\ref{fig:leaktest}(c)). O-ring seals are lubricated
with IPA to prevent twisted o-rings.  IPA is used as a lubricant since it completely evaporates and leaves no residue that might contaminate the drift gas.
The tube is inserted into the vessel, 
the end flange is sealed, and the leak detector set to pump out the vessel.
When the background He level reads about 
$\rm 1\times10^{-6} cm^3\times mbar/s$ the tube is filled with Helium gas 
to a pressure of 3 barA. 
Leaks can be found within 30 seconds of pump-out.
When this test is completed the Helium is vented through a snorkel to the outdoors.
\par
The goal of this test step is to establish the tightness of the seal.
No data was stored for this test, only pass/fail. 
All the tubes built at UM passed the ATLAS gas tightness requirement.

%% file: tube_length_wire_tension.tex
\subsection{Tube length and wire tension measurements}
\label{sec:WireTension}

\begin{figure}[ht]
  \centering
  \includegraphics[width=0.95\linewidth]{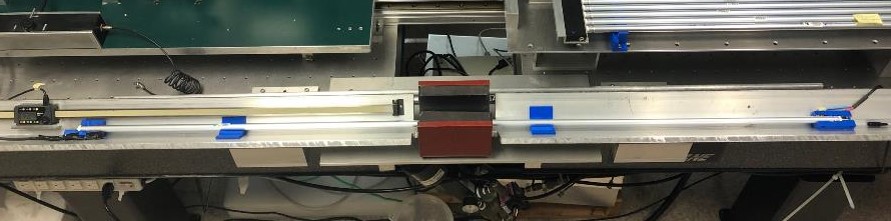}\\
  \caption{Test Station for tube length and wire tension measurements.
  }
  \label{fig:LengthSetup}
\end{figure}

\begin{figure}[!hbt]
    \centering
    \includegraphics[width=0.65\linewidth]{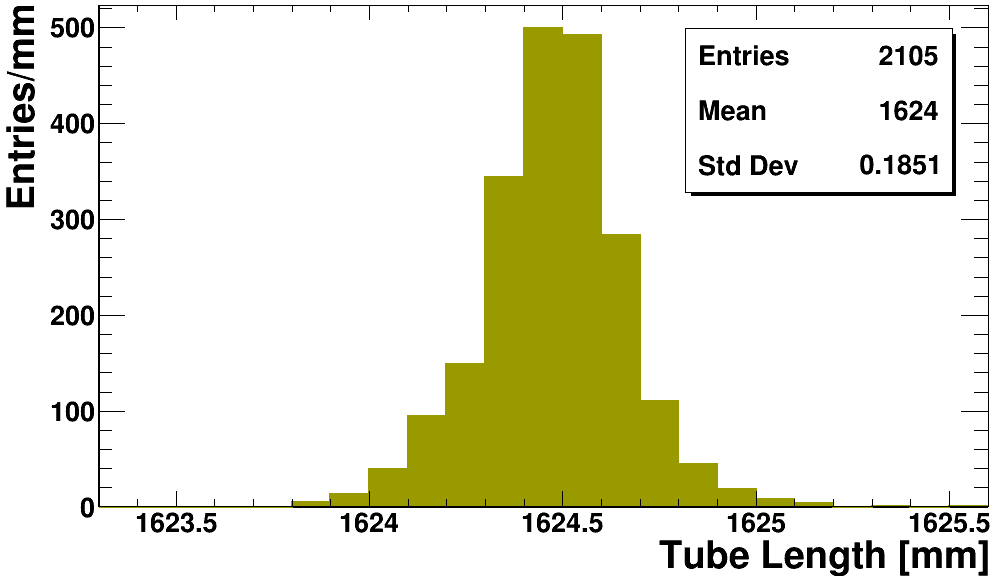}
    \caption{Distribution of the measured length 
    of the tubes built at UM.}
    \label{fig:LengthResult}
\end{figure}

The tube length and the wire tension are measured 
using the station shown in Figure~\ref{fig:LengthSetup}.
A tube is placed in the V-shaped aluminum bar with its midpoint inside a U-shape magnet, and a linear digital scale~\footnote{Mitutoyo ABSOLUTE Digimatic Scale Units Series 572, https://www.mitutoyo.com/webfoo/wp-content/uploads/E316-572R\_ABS\_Digi\_ScaleUnits.pdf}
is used for the length measurement. 
Before any session of measurements, the  digital linear gauge (precise to 
$ {} \pm 10 \; \mu$m) is calibrated  with an 
aluminum rod of length $ 1624.5 \pm 0.1 $~mm.
Then a tube is placed into a V-shaped bar with 
the two ends resting in 3D printed saddles 
(in blue in Figure~\ref{fig:LengthSetup}) 
so that one end can be pressed against a stopper.
The operator makes sure that a good electrical 
contact is made at both extremities (tube
grounded and wire connected the electrical circuit).
The digital scale head is brought into contact 
with the endplug of the tube and the length of
the (fully assembled) tube is recorded.
The measured length distribution of the tubes 
built at UM  is shown in 
Figure~\ref{fig:LengthResult}.
A maximum difference in tube length slightly 
larger than 1~mm  is observed. 
None of the tubes built at UM were rejected 
due the length being out of specification.

%Electric signals generated by a custom circuit 
%(see Figure~\ref{fig:circuit}) trigger vibrations in the wire and
%measurements of the wire vibration spectrum are made
%through the electrical connections at the two ends 
%of the tube and read into a computer.
%The wire vibration frequencies are measured 
%through a Fourier transformation of the wire 
%vibration spectrum.
\par
\begin{figure}[hbt]
  \centering
  \includegraphics[width=0.99\linewidth]{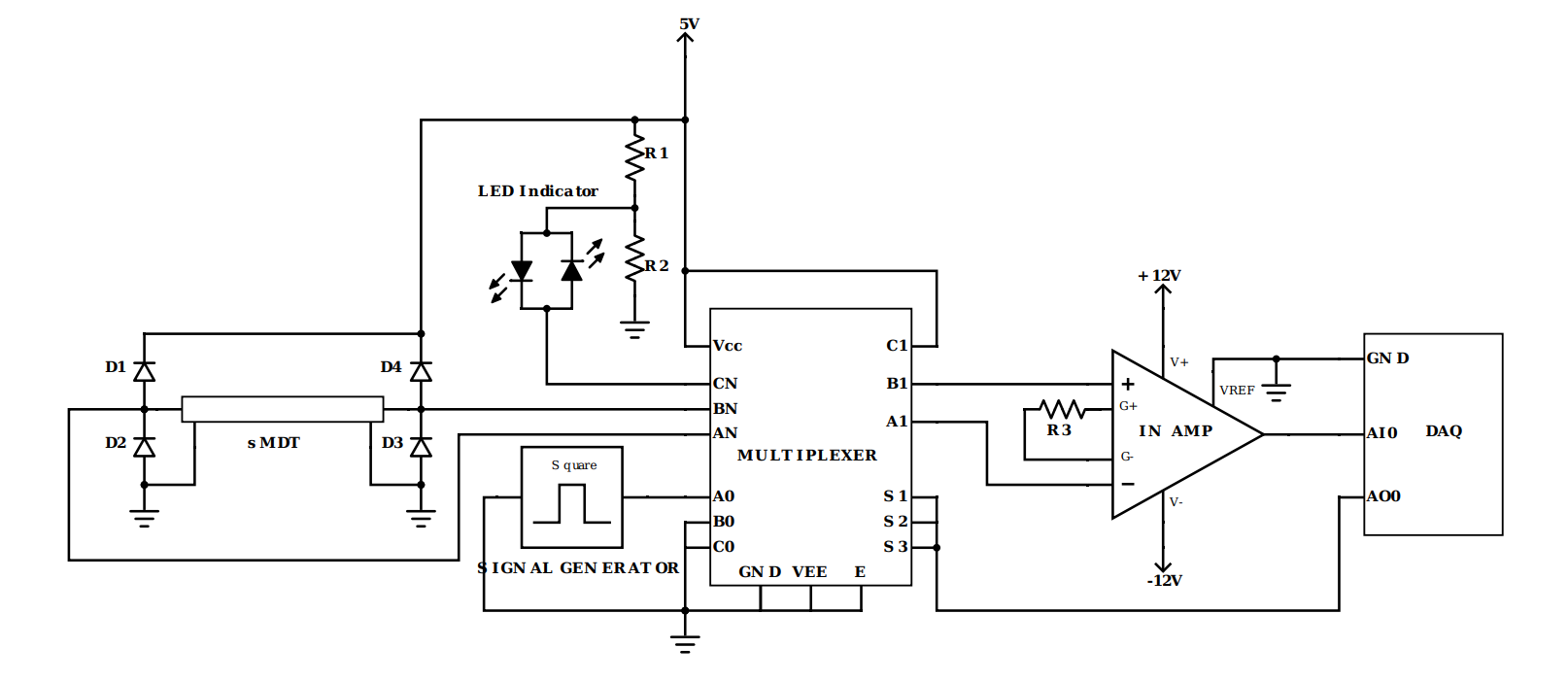}
  \caption{Circuit for wire tension test.
  }
  \label{fig:circuit}
\end{figure}

Wire tension is then measured using the same set up with the following procedure.
The aluminum tube is tied to the ground on the V-shaped bar and the wire is connected to the  custom multiplexing circuit shown in Figure~\ref{fig:circuit}.
The circuit consists of an analog multiplexer, signal generator,
instrumentation amplifier, and the sMDT itself. The role of the multiplexing circuit is to trigger wire vibrations in the sMDT by directing pulses from the signal generator to the sMDT and then to send the signal
from the discharging sMDT to the instrumentation amplifier. 
The instrumentation amplifier is connected to a data acquisition (DAQ) system which consists of an multipurpose I/O device (NI USB-6008\footnote{National Instruments USB-6008 Multifunction I/O Device, https://www.ni.com/en-us/support/model.usb-6008.html}) interfaced to a computer.
A software program with a graphic user interface was developed
to allow setting of wire tension measurement parameters, such as the measured tube length, and to perform real time analysis and measurements.
%
%In this same test stand a current signal is sent 
%through the wire, inducing a mechanical vibration 
%of the wire which is in a static magnetic field.

After the external pulse is stopped the wire 
oscillates at its natural frequencies 
(see eq. \ref{eq:tension} from~\cite{TensionPaper}):
\begin{equation}
  f_{n}^{} = \frac{n}{2L} \sqrt{\frac{T}{\rho}},
  \label{eq:tension}
\end{equation}
that converted to the tension expressed in 
grams is $  T =  \pi L^2 d^2 f_{1}^2 \rho / g $.
With the nominal sMDT wire parameters 
$\rm L = 1597\;mm$, 
$\rm d=50 \; \mu m$,
$\rm \rho = 19.7\; g/cm^3$ and $\rm g = 9.81 \; m/s^2$,
the first resonant frequency of 93.3~Hz 
corresponds to a tension of 350~g.
The frequency is measured with a custom LabView program 
which uses a fast Fourier transform (FFT) to 
find the first resonant frequency and thus the wire tension.
\par

%The electric signals which trigger wire vibrations are generated by the custom circuit shown in Figure~\ref{fig:circuit}. The mechanical oscillations drive a corresponding  oscillating current in the wire which is read by  a data acquisition module (NI USB-6008). 
%This is analysed by a custom LabView program 
%which uses a fast Fourier transform (FFT) to 
%find the first resonant
%frequency and thus the wire tension.

%where $f$ is the frequency of oscillations,
%$L$ and $d$ are the wire length and diameter,
%$\rho$ is the wire linear density (for a circular wire 
%of density $\varrho$, $\rho=\varrho/(\pi\cdot r^2$)),
%$T$ is the tension (in grams) stretching the wire,
%$n$ is the order of the harmonic, and
%$g $ is the gravitational acceleration (used here to
%quote the tension in grams instead of the Newtons).
%\par
%The frequencies of these oscillations are measured 
%from the waveform of the induced Faraday current by a 
%NI~USB-6008~\cite{NI6008} with an accuracy up to 0.2~Hz,
%leading to an error on the tension of ${} \pm 0.9 $~g.
%Tubes with wire tension in a range, $\rm 350 \pm 20 \; g $ are accepted in the sMDT mass production.
%\par
%The whole measurement is automated into a LabView
%custom program analyzing the Fourier transform (FFT) 
%of the signal waveform determining the frequency of 
%the oscillations  and thus the wire tension.
In order to have a standard definition of tension 
and compare the results measured at UM with those 
measured by MSU during the tube construction, 
the value quoted uses the nominal (fixed) length 
of 1597~mm, so that only the frequency ($f_1^{}$) 
is a measured parameter.
\par
A second tension measurement is made at least two 
weeks after the first measurement to check for any 
significant drop in tension. 
If a drop larger than 18~g in 2~weeks is found, the
tube is rejected.
\par
\begin{figure}[ht]
    \centering
%    \includegraphics[width=0.49\linewidth]{images/UM_HZ.png}
%    \includegraphics[width=0.49\linewidth]{images/UM_TEN.png}
%    \caption{Tube wire first harmonic frequency (left) and  
%             tension (right) measurements.}
        \subfloat[]{
    \includegraphics[width=0.49\linewidth]{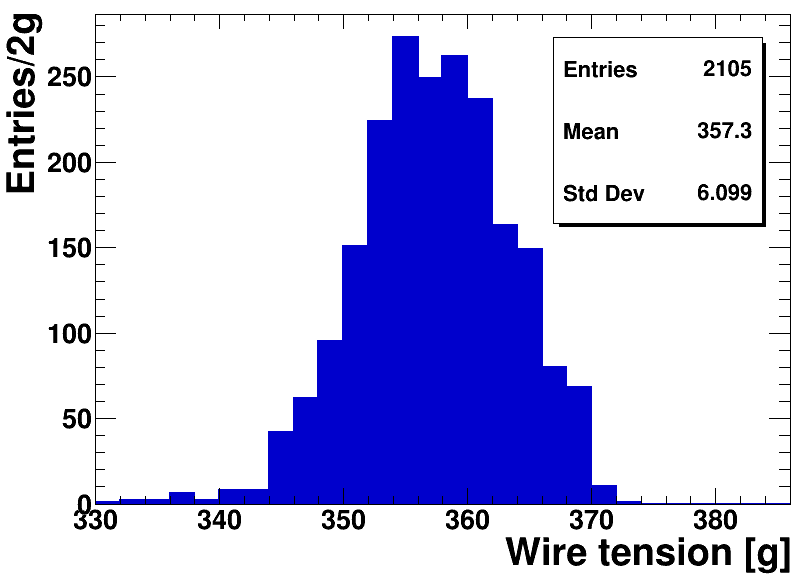}
    }
            \subfloat[]{
    \includegraphics[width=0.49\linewidth]{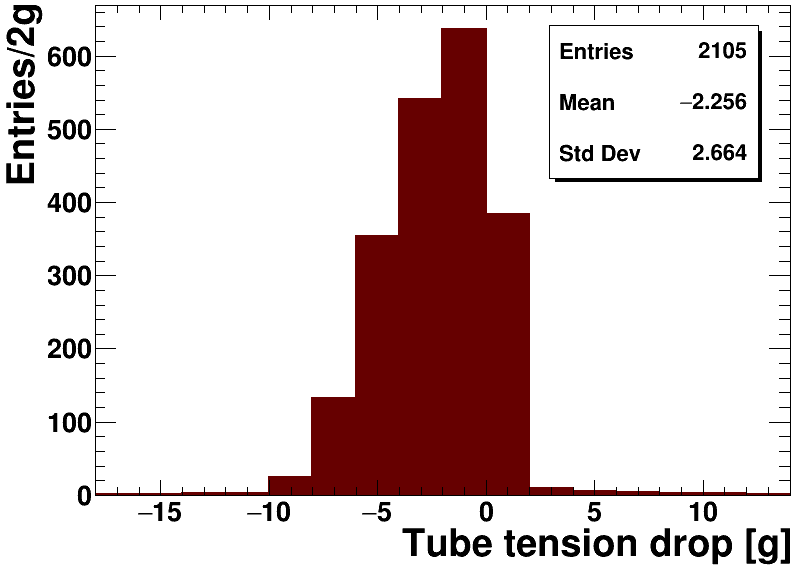}
    }
    \caption{(a) Tube wire tension, 
             and (b) tension drop after at least 2 weeks.}
    \label{fig:TensionResult}
\end{figure}

The results of the wire tension measurements for 
tubes made at UM are shown in 
Figure~\ref{fig:TensionResult}. 
A total of 2 tubes out of the 2105 built (0.09\%) 
did not pass the tension tests.
\\

%\begin{figure}[ht]
%    \centering
%   \includegraphics[width=0.55\linewidth]{images/UM_TE2.png}
%   \caption{Tube wire tension difference after at least 2 weeks.}
%    \label{fig:Tension2Result}
%\end{figure}

%% file: high_voltage_test.tex
\subsection{Dark current measurement}
\begin{figure}[htbp]
    \centering
    \includegraphics[width=0.9\linewidth]{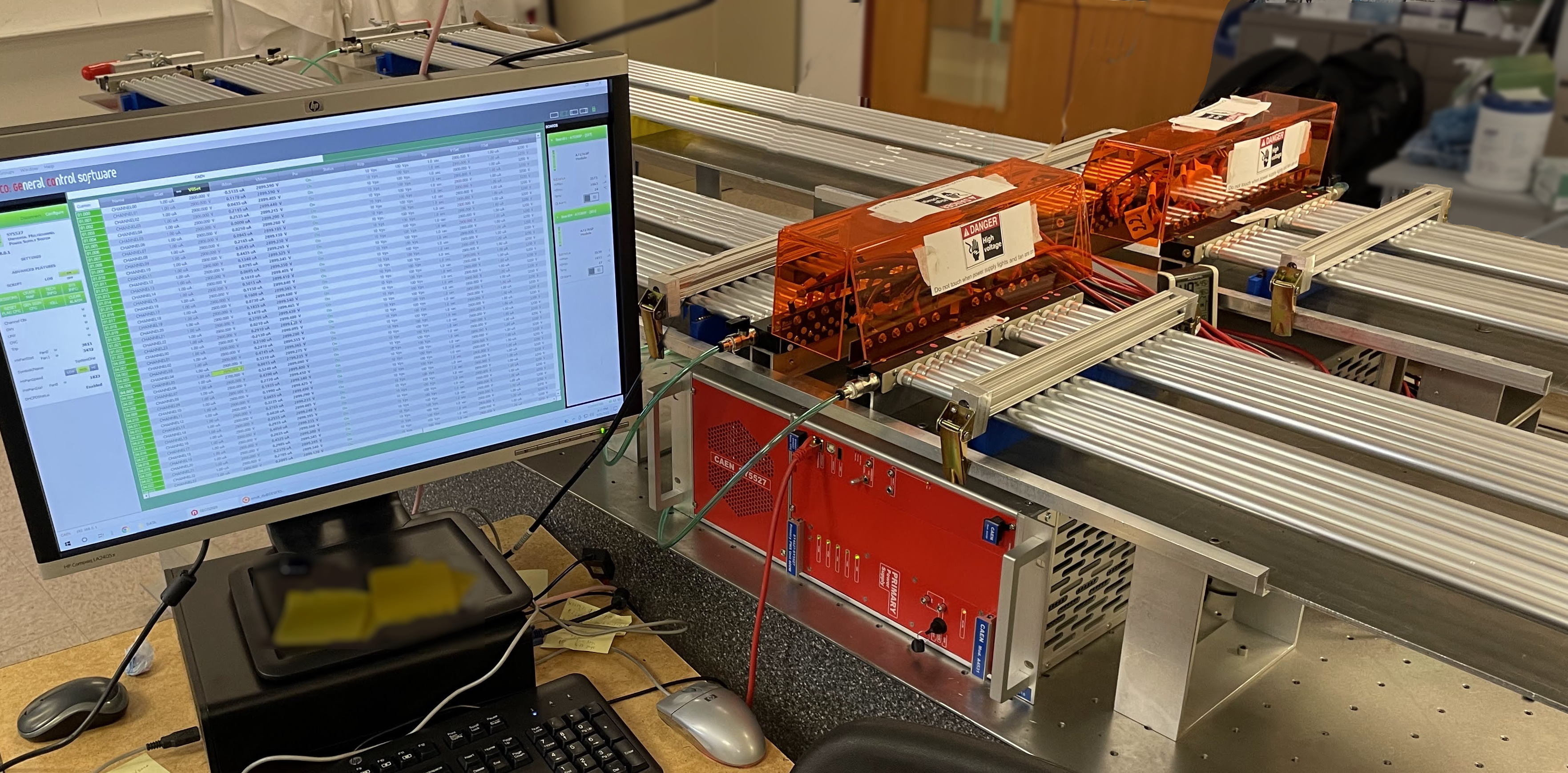}
    \caption{Test Stand with 48 tubes with the CAEN  
    SY5527 mainframe, with the CAEN GECO monitoring 
    program running in the connected computer.}
    \label{fig:CAENmounting}
\end{figure}

The tube, at 3 atmospheres absolute pressure of the 
working gas mixture, Ar:CO$_2$ (93:7), operated at 2800 volts 
and must draw less than 2~nA of dark current.
The high voltage (HV) test set up (shown in Fig.~\ref{fig:CAENmounting}) 
and procedure are described in this section.

\subsubsection{HV test station}
\label{sec:TubeMounting}
%Tubes built and dark-current tested at MSU are re-tested for dark current at UM. 

\begin{figure}[htbp]
    \centering
    \includegraphics[width=0.95\linewidth]{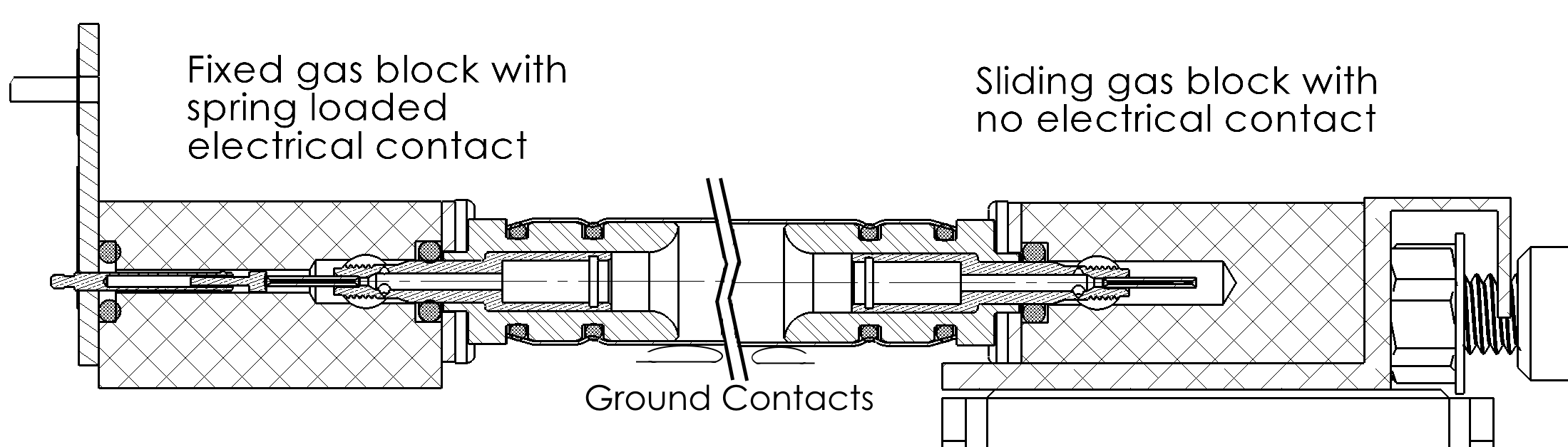}
    \caption{Tube mounting diagram for dark current test station}
    \label{fig:DCTestXSection}
\end{figure}

The HV test station at UM holds 4 groups of 12 tubes.
Tubes are inserted into 4 sets of gas-manifold blocks to pressurize 
them. The gas-manifold blocks, machined from acytal plastic
(polyoxymethylene POM), allow 12 tubes to be plugged in with a
simple push of a handle to make the tube mounting process simple and quick. The 5~mm cylindrical precision brass part of the 
endplugs of the 12 tubes are inserted into thick o-rings which 
are built inside the gas-manifold to make a gas-tight seal.
The design drawing of the set up is shown in 
Figure~\ref{fig:DCTestXSection}.

%\subsubsection{Gas System for HV Test} 
%\label{sec:GasSystem}
The input gas pressure is regulated to 3~barA and fanned out to the 4 gas distribution
blocks, each with
its own input and output shutoff valves. 
The output of the 4 blocks is fanned into a 
single flow meter and a bubbler. 
Initially the flow is set high (3 SCFH) 
for an hour to flush at least 3 volumes through 
the tubes. 
It is then turned down to a few bubbles per 
second (flowing around one gas volume per day) for testing.

\subsubsection{Power supply system}

The power supply system consists of a CAEN SY5527~\footnote{Universal
Multi-channel Power Supply System with 4 Channel 
3.5~kV, 1.5/0.15 mA (4W) Common Ground Dual Range 
Boards, https://www.caen.it/products/sy5527/} 
mainframe which hosts two 24-channel AG7326 HV 
modules, able to deliver up to 3.5 kV per channel
with 500~pA resolution for current measurements,
with GECO (GEneral COntrol) monitoring and data
acquisition software.
Two places along the tubes are held rigidly to make good contact 
with a grounding strip.  % of finger stock.
One of the two end manifolds is fitted with spring-loaded 
contacts for applying high voltage to the wire. 
The spring contacts are hard-wired to HV cables plugged
into 48 individual channels of two modules of the CAEN HV 
power supply.
%
%\begin{figure}[htbp]
 %   \centering
%    \includegraphics[width=0.50\linewidth]{images/CAENmainframe.png}
%    \includegraphics[width=0.40\linewidth]{images/UMtubesAndPS.jpg}
%    \caption{CAEN mainframe (left) and custom Power 
%    Distribution system for the UM station (right).}
%    \label{fig:PowerSource}
%\end{figure}
%
For a correct dark current measurement, the
pedestal current level for each channel must be subtracted 
from the current measurement.  The pedestals are found by
turning on the high voltage with no tubes connected and measuring the average current on each channel for a period of one hour.
The pedestals are stable except if the humidity becomes too high and must be re-calibrated if the relative humidity exceeds 50\%.

\subsubsection{Dark current measurement}
\label{sec:DC-test}

\begin{figure}[htbp]
    \centering
            \subfloat[]{
    \includegraphics[width=0.48\linewidth]{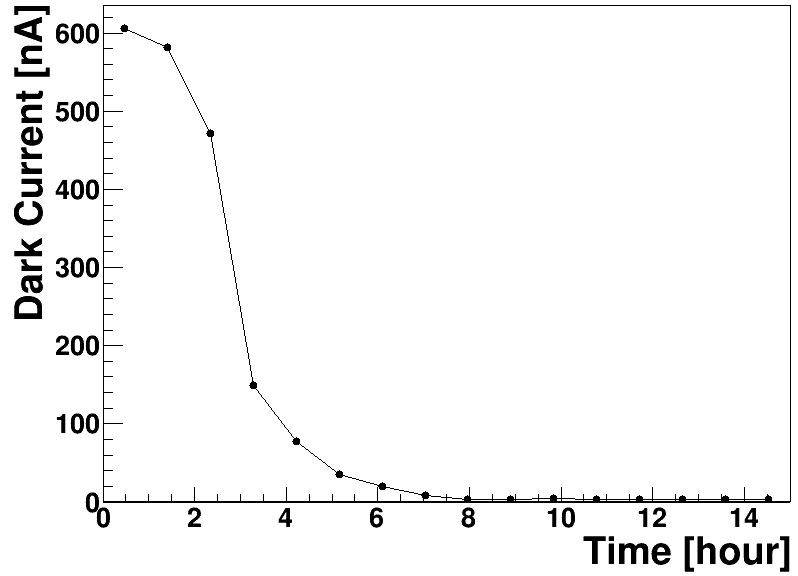}
    }
            \subfloat[]{
    \includegraphics[width=0.48\linewidth]{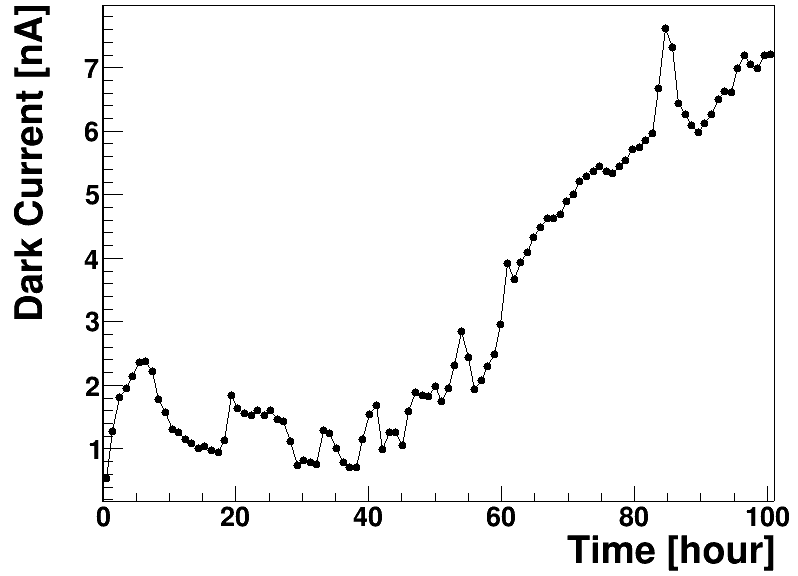}
    }
    \caption{(a) Example of a tube recovered from initial high 
    dark current, and (b) one that instead developed 
    higher and higher dark current in time. Note the different scales.}
    \label{fig:BadTubes}
\end{figure}

To measure the dark current the tubes are placed in the gas manifolds
and the voltage is ramped up to +2,900 V and dark current readings are  
are recorded for at least 4 hours. If the average current
over the last hour of the test is below 2~nA, the tube
passes the test.
Some tubes show dark current above the limit in the
first hours after the HV was turned on, very often with a
decreasing trend as shown in  Figure~\ref{fig:BadTubes} (a),
but sometimes the dark current increases in time leading to 
a tube failing the test (see Figure~\ref{fig:BadTubes} (b)).
%Foreign material (e.g. dust) inside the tube may stick to 
%the wire and cause discharge. This can be burned off over time 
%(and the residues removed from the gas volume) resulting in a  
%reduction in the dark current to an acceptable level.
Dark current are retested for all tubes received from MSU 
to insure they are fully burned-in before being used in 
chamber construction.

\subsubsection{Tube burn-in with negative HV}

Tubes that failed the dark current test are treated 
with negative HV (-3~kV for ~30--60 minutes) on a separate test 
station, and then reassessed via the standard HV test.
If the tube fails again, then treatment with negative HV is repeated.
All tubes with high dark current have responded to this 
treatment so that not a single tube had to be rejected 
in the dark current test and treatment procedure.
\par
\begin{figure}[htbp]
    \centering
            \subfloat[]{
    \includegraphics[width=0.48\linewidth]{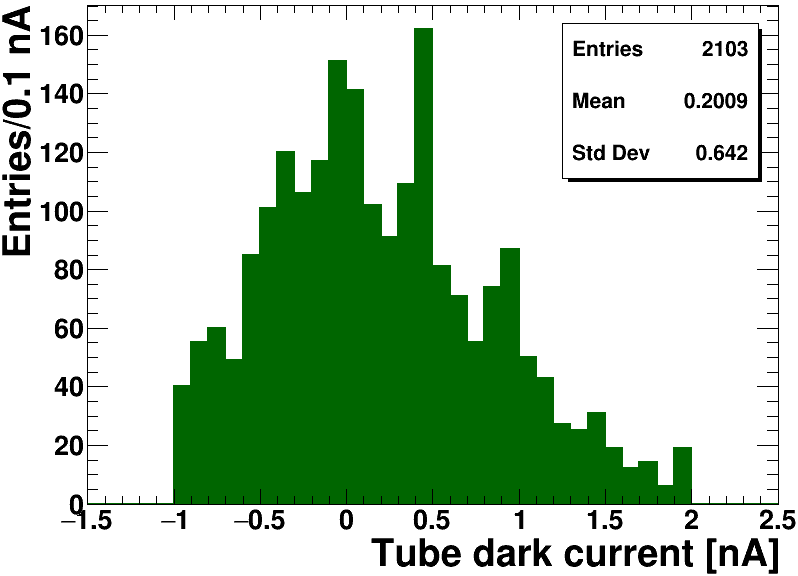}
    }
            \subfloat[]{
    \includegraphics[width=0.48\linewidth]{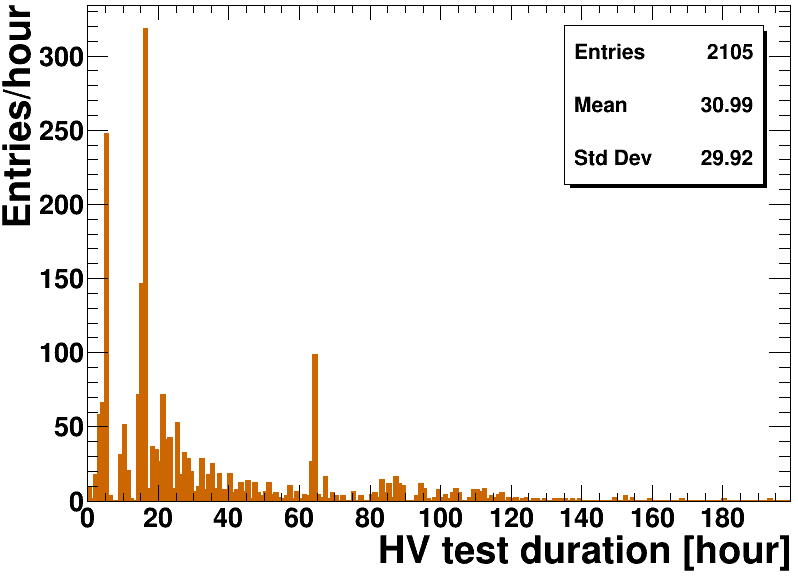}
    }
    \caption{(a) Distribution of the measured final tube dark current, and (b) the burn-in time (right), where the three spikes corresponds to the most common test lengths: day time (just a few hours), overnight (typically about 15-16 hours) and over the weekend (normally around 64 hours).}
    \label{fig:DarkCurrent}
\end{figure}

In the 2105 tubes built at UM in the summer of 2021, 
a total of 67 tubes (3.2\%) were initially measured 
to have dark current above 2~nA.
All of them were recovered with either longer HV burn-in time or via the negative HV treatment.
Figure~\ref{fig:DarkCurrent} (a) and (b) show the measured dark current distributions of the UM built tubes, and the tube burn-in time distributions, respectively.

%% file: tube_database.tex
\section{Tube Production Database}

All the information gathered in tube 
production and testing is recorded into a local
database (DB) for production QA/QC monitoring, and a subset of the data
is uploaded to a master chamber production DB encompassing both US and German chambers.
UM receives the QA measurements for the 
tubes delivered by MSU, used in part as a first 
data-point in the history of the tubes tests at 
UM, and the UM results are shared back with MSU 
for cross-checks and validation.
All the test results from the UM test stations are 
saved in text files, with a format that can be read by 
different databases 
(MS Access, MySQL, Oracle, Jason, etc).
The most natural choice is a relational database, 
since all information to be saved is relative
to many single tubes, each one with a history 
of production and test results.  In the tube DBs the barcode serial 
number is used as the primary key for retrieving  information. 
The barcode is also the link between the entries of the 
tubes layout in a sMDT chamber and the associated 
test results.
\par
ROOT~\cite{root}, an analysis framework used commonly in high 
energy physics, is chosen locally to store, 
retrieve, and quickly plot the distributions 
of the test results. 
A few C++ classes have been developed to read the 
station measurement results and convert them 
into a set of text files for each tube. 
At the end of the process all the test 
conditions and data relative to a tube are 
saved into a "TTree" of the ROOT file with 
an entry for each tube. 
%
%An example for the wire tension station is shown in Fig.\ref{fig:smdtTT}: on the right figure the first seven quantities are single values (including the results of the second tension test), "ntens" is the counter of the number of tension tests performed, and all others are arrays of parameters with as many elements as number of tests.
%\begin{figure}[htbp]
%    \centering
%    \includegraphics[width=0.2\linewidth]{images/smdtTT_Tree.png}
%    \hspace{2 cm}
%    \includegraphics[width=0.45\linewidth]{images/smdtTT_Header.png}
%    \caption{Example of the ROOT TTree "smdtTT" storing 
%    the wire tension test results (left) and its C++ 
%    class structure used for accessing it (right).}
%    \label{fig:smdtTT}
%\end{figure}\\
%
To ease access to tube test data, separate "TTrees" have been created for 
each station. In addition, there is a "master" tree 
with all the tube test information. 

\par
While filling the "TTree", a set of standard plots 
for each test is automatically saved and a local 
web page is automatically created to show the 
results history of all tests. 
%(see fig.\ref{fig:smdtWeb}), 
%with a pull down menu to select the desired tube that opens a new 
%frame with distributions of results for every test station.
\par
Eventually, all the production and test 
information are uploaded to the CERN sMDT production 
database which records information for all of the detectors 
to be installed in the ATLAS experiment.
%
%\begin{figure}[htbp]
%    \centering
%    \includegraphics[width=0.5\linewidth]{images/smdtWEB.png}
%    \includegraphics[width=0.4\linewidth]{images/smdtWEB_MSU00245.png}
%    \caption{Web page for the overall tube result 
%    distributions with pull-down menu to single tubes 
%    (left) and example of one (out of five) summary plot 
%    for a tube (right).}
%    \label{fig:smdtWeb}
%\end{figure}

%% file: conclusion.tex
\section{Conclusions}

Fully functional stations for sMDT tube production and testing
were designed, built, and successfully used at the University of Michigan 
for the ATLAS HL-LHC muon detector upgrade project.
A total of 2105 tubes were built at UM and tested for sMDT 
chamber production. The rejection rate of the UM constructed 
tubes is below 0.1\% from the QA tests.
In addition, over 14,000 tubes built at MSU have been fully 
tested at UM.
\par
Tubes passing all tests have been used to construct 30 sMDT 
chambers so far out of the 50 of the whole production task. 
All the chambers meet the stringent precision and performance
requirements, which will be reported in a separate paper.
%
%At the end of the tube certification 2 out of 2105 tubes built at UM were rejected (less than 0.1~\%) because of the wire tension outside the nominal range of $ 350 \pm 20 $ grams and none due to the dark current above the limit of 2~nA.

\section*{Acknowledgements}
We would like to thank Hubert Kroha~(MPI) and Reinhard Schwienhorst~(MSU) and their team members
for their close collaborations and valuable technical discussions;
Toni Baroncelli (USTC) and Massimo Corradi (INFN Roma) for coordinating 
the contract with MIFA for aluminum tube production and delivery; 
and Rinat Fakhrutdinov (IHEP, Protvino) for producing the endplugs 
and associated components. 
%the engineers at MPI and at MSU for the drawings of parts and the collaborative exchange of information.
We also thank the UM faculty, Shawn McKee, Tom Schwarz, Junjie Zhu, and Jianming Qian, for their strong support; the UM electrical engineers,
Yuxiang Guo,  Xiangting Meng, and Jinhong Wang, and former graduate student Christopher Grud, for their help in the initial R\&D phase to 
design and build the required electronics; 
%and development of the online DAQ programs for tube wiring control and testing; 
and the UM undergraduate students, Tyler Coates,  Greg Kondas, Kathryn Ream, and Noah West, for their great technical assistance.
%
%Special thanks to Xiangting Meng (former UM post-doctoral fellow) and Christopher Grud (former UM graduate student) for their great efforts in developing the electronic circuits and online control programs for tube wiring and tension test. We also thank Jinhong Wang (former UM electrical engineer) for the design of the PCB for tube dark current test.
\par
This work was supported in part by the U.S. Department of Energy grant
DE-SC0012704 %(BNL assigned 326127 for UM) 
and in part from the U.S. 
National Science Foundation PHY -1948993. %(Columbia assigned 17(GG016228) for UM).